# Using a Bessel light beam as an ultra-short period helical undulator

B. C. Jiang, Q. L. Zhang and Z. T. Zhao

Shanghai Institute of Applied Physics, Chinese Academy of Sciences, Shanghai 201800, China

The undulator is a critical component to produce synchrotron radiation and free electron laser. When a Bessel light beam carrying the orbit angular momentum co-propagates with an electron beam, a net transverse deflection force will be subjected to the electron beam. As a result of dephasing effect, the deflection force will oscillate acting as an undulator. For such a laser based undulator, the period length can reach sub-millimeter level, which will greatly reduce the electron energy for the required X-ray production.

1. **Introduction**

The magnetostatic undulator is made of periodic structures of dipole magnets [1,2]. The static magnetic field of the undulator is perpendicular to the electron beam trajectory and periodically changes its directions, which causing the electrons follows an undulating trajectory to radiate energy. The brightness of the radiation from the undulator at some wavelength is $N^2$ times higher than that from a single bending magnet, where $N$ is the total period number of the undulator.

The relationship of the radiation wavelength to the undulator period length is as follows[3]:

$$\lambda_{rad} = \frac{\lambda_u}{2\gamma^2}(1 + \frac{K^2}{2} + \gamma^2\theta^2), \qquad (1)$$

$$K = \frac{eB\lambda_u}{2\pi m_e c} = 0.934 B(T)\, \lambda_u(cm), \qquad (2)$$

Where $\lambda_u$ is the period length of the undulator, $B$ is the peak magnetic field of the undulator, $\lambda_{rad}$ is the radiation wavelength, $\theta$ is the radiation angle. $\gamma, e, m_e$ are the Lorentz factor, charge and the rest mass of the electron respectively, $c$ is the speed of light.

An important development direction of the undualtor is to decrease the period length. The shorter period length of the undulator, the lower electron energy required for a desired X-ray radiation, which may greatly reduce the scales and the costs of the facility.

For a practical configuration, $K$ value of the undulator should be on order of 1. To this end, shorter the period length, higher the peak magnetic field should be, which prevents the magnetostatic undulator period to reach ultra-short. In-vacuum undulator is developed for short period approach [4,5]. For this type of undulator, the permanent magnets are exposing to the electron beam in the vacuum tank thus the undulator poles gap can be much smaller. The peak field is increased and then the period length can be reduced. The discovery of increasing remanent field and coer-civity of the permanent magnet at low temperature makes the cryogenic permanent magnet undulator (CPMU) [6,7] possible to achieve a little bit shorter period at the costs of an additional liquid nitrogen cryogenic system. Taking the advantages of the superconducting technology, even shorter period undulator is available [8,9]. However, even for the technology of state of the arts, the magnetostatic undulator period length is beyond 1mm[10]. RF undulator [11, 12] is possible to achieve period length shorter than 1mm. However, for lacking of high power THz source, millimeter period undulator is still hard to reach.

Optical undulator has been proposed for compact FEL purpose for years. When an intense and long enough laser pulse counter-propagate with the electron beam, the laser may act as an undulator[13, 14, 15]. The period length of optical undulator is in micron range which requires the electron beam orders of magnitude brighter than the existing electron source for FEL producing. Laser plasma undulator [16, 17] has recently been proposed to produce sub-millimeter period undulator. The electron is not passing through free space, which may prevent it be used in storage rings.

In another area, the light beam carrying orbital angular momentum (OAM) was unfamiliar until it is discovered in Laguerren-Gaussian beams [18,19]. Since its discovery, OAM has attracted myriad modern interests. Among them, trapping and accelerating charged particles is an amazing application [20, 21, 22, 23]. Bessel beam, a more simple solution of the light wave in open free space carries OAM, for which, the field can be separated into transverse and longitudinal parts. Bessel beam is usually used to analyze trapping of charged particles. In the study of trapping charged particles, Bessel beam has already shown the ability of wiggling the charged particles in low energy cases, while for the interaction with the relativistic electron, its properties is still unveiled.

In the following sections we will show that Bessel light beam can also be used for undulating the relativistic electrons when they are co-propagated as shown in Figure 1. With the advent of the high power laser, it is possible to achieve *K* value of the Bessel light beam undulator (BLU) around 1. And the undulator period length can approach sub-millimeter.

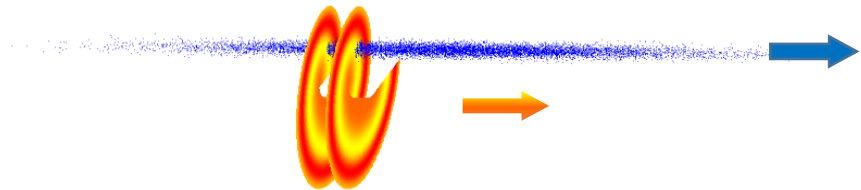

Figure 1. Sketch of Bessel light beam (red vortex) interaction with electron beam (blue dot)

## 2. Transverse force of the Bessel light beam

For a monochromatic Bessel light beam, in the dimensionless system (*c=1*), the electric and magnetic fields of the wave in Cartesian coordinates for paraxial approximation can be expressed [24]:

$$\begin{pmatrix} \mathcal{E}_x \\ \mathcal{E}_y \\ \mathcal{E}_z \end{pmatrix} = \begin{pmatrix} \kappa_- C_{M+1} + \kappa_+ C_{M-1} \\ \kappa_- S_{M+1} - \kappa_+ S_{M-1} \\ 2 S_M \end{pmatrix}, \quad (3\text{-a})$$

$$\begin{pmatrix} \mathcal{B}_x \\ \mathcal{B}_y \\ \mathcal{B}_z \end{pmatrix} = \begin{pmatrix} \kappa_- S_{M+1} + \kappa_+ S_{M-1} \\ -\kappa_- C_{M+1} + \kappa_+ C_{M-1} \\ 2 C_M \end{pmatrix}, \quad (3\text{-b})$$

$$C_M = cos(k_\parallel z - \chi \omega t + M\phi) J_M(k_\perp \rho), \quad (4\text{-a})$$
$$S_M = sin(k_\parallel z - \chi \omega t + M\phi) J_M(k_\perp \rho), \quad (4\text{-b})$$

where $z$ is the light propagation direction, $\rho = \sqrt{x^2+y^2}$ is the transverse distance to the $z$ axis and $\phi$ is the azimuthal phase to the $z$ axis. $M$ is the order of Bessel beam, where

$$\kappa_\pm = \frac{k \pm k_\parallel}{k_\perp}. \quad (5)$$

In this paper only the forward propagated wave is considered which is $\chi = 1$. The wavenumbers have the following relationships,

$$k = \sqrt{k_\parallel^2 + k_\perp^2}, \quad (6)$$
$$\omega = c\,k, \quad (7)$$

The Lorentz force in the horizontal ($x$) and the vertical ($y$) direction for a co-propagated electron can be derived:

$$F_x(z,t,\rho,\phi) = -e(\mathcal{E}_x - c\,\mathcal{B}_y) = -2e\kappa_-\,C_{M+1}(z,t,\rho,\phi), \quad (8\text{-a})$$

$$F_y(z,t,\rho,\phi) = -e(\mathcal{E}_y + c\,\mathcal{B}_x) = -2e\kappa_-\,S_{M+1}(z,t,\rho,\phi). \quad (8\text{-b})$$

Forces in $x$ and $y$ plane expressed in Eq.(8) oscillate in a *sin* waveform with phase difference $\pi/2$, forming a force like a helical undulator.

From Eq.(4,6,7), it can be find out that the phase velocity of the Bessel beam is faster than the speed of light. Assuming the relativistic electron get the velocity very close to the speed of light (v ≅ c), the phase slip of a relativistic electron to the light produces upconversion undulate period of the laser wavelength which is:

$$\lambda_u = \frac{1}{1-k_\parallel/k} \lambda_{laser}. \quad (9)$$

The key factor of a Bessel light could undulate the relativistic electron is that its phase velocity is faster than the speed of light. The magnetic field can't be cancelled completely by the electric field when seen by a co-propagated relativistic electron, leaving a net periodic oscillated deflecting force.

### 3. Determinants of the undulator deflection parameter *K*

In this section we will analyze the undualtor deflection parameter *K*. In the case of $k_\perp \ll k$, it can be found out $\kappa_+ \gg \kappa_-$. The transverse EM field expressed in right hand of Eq. (3) is then mainly composed by the term with coefficient $\kappa_+$ such as:

$$\mathcal{B}_y \approx \kappa_+ C_{M-1}. \quad (10)$$

Assuming the transverse size of the electron beam $\sigma \ll 2\pi/k_\perp$, the electron beam travels at a certain transverse place $(\rho_0, \phi_0)$ in the Bessel light beam, following relationship can be written:

$$\widehat{\mathcal{B}_y(z,t)}|_{\rho_0,\phi_0} \approx \kappa_+ \widehat{C_{M-1}(z,t)}|_{\rho_0,\phi_0}, \quad (11)$$

where $\widehat{B}$ denotes maximum value of the B.

The transverse force in horizontal plane can be written as:

$$F_x(z,t)|_{\rho_0,\phi_0} = -2e\frac{\kappa_-}{\kappa_+} C_{M+1}(z,t)|_{\rho_0,\phi_0}\kappa_+ \approx -e(1-\frac{k_\parallel}{k})C_{M+1}(z,t)|_{\rho_0,\phi_0}\kappa_+. \quad (12)$$

Substitute Eq.(11) to Eq.(12) we get:

$$F_x(z,t) \approx -e(1-\frac{k_\parallel}{k})\frac{\widehat{\mathcal{B}_y(z,t)}}{\widehat{C_{M-1}(z,t)}}C_{M+1}(z,t). \quad (13)$$

The force of the laser is equivalent to a static magnet with peak value:

$$\widehat{B_{y\_eff}} \approx (1-\frac{k_\parallel}{k})\frac{\widehat{C_{M+1}(z,t)}}{\widehat{C_{M-1}(z,t)}}\widehat{\mathcal{B}_y(z,t)}. \quad (14)$$

Thus the undulator defection parameter $K_x$ is:

$$K_x = 0.934\widehat{B_{y\_eff}}\lambda_u = 0.934 \times \frac{\widehat{C_{M+1}(z,t)}}{\widehat{C_{M-1}(z,t)}}\widehat{\mathcal{B}_y(z,t)}\lambda_{laser}. \quad (15)$$

In vertical plane, it can be treated in the same way and is not repeated here.
For circular polarized helical undulator the total defection parameter is:

$$K = \sqrt{2}K_x. \quad (16)$$

The expression of parameter $K$ of the BLU gets no factor of $k_\perp$, only determined by the laser power density, the wave length and the Bessel light beam order $M$. It is consistent with the RF and optical undulator. When degenerate Bessel beam to a plane wave laser ($k_\perp = 0$), as it co-propagate with the electron, $E$ field is almost canceled by the $B$ field, the Lorentz force is $B_{eff} = (1-\beta)B_y$, while the electron dephasing in the light makes $\lambda_u = \lambda_{laser}/(1-\beta)$. Where $\beta$ is the electron velocity in units of light speed. As the plane wave counter-propagate with the electron, $E$ field add up with the $B$ field, $B_{eff} = 2B_y$, and the undulator period $\lambda_u = \lambda_{laser}/2$. For both of above two cases $K = 0.934B_{eff}\lambda_u = 0.934\widehat{\mathcal{B}_y}\lambda_{laser}$.

4. **Laser power requirement**

The ideal Bessel light beam has an infinitely extended transverse profile and carries infinite power. Bessel light beams created experimentally are non-ideal ones with finite radius and finite power. A non-ideal Bessel beam with spot radius $R$ gets diffract distance [17]:

$$L = R\frac{k}{k_\perp}, \quad (17)$$

The Bessel light beam holds $N$ rings within radius $R$ will be diffracted layer by layer until the innermost ring diffracts away at the end of the diffract distance.

For $\frac{k_\perp}{k} \ll 1$, the laser power is mainly determined by the electric and magnetic term in right hand of Eq.(3) with $\kappa_+$ coefficient. The power can be integrated approximate as follows:

$$P \approx 2\kappa_+^2 \int_0^{2\pi} d\phi \int_0^R \rho J_{M-1}(k_\perp \rho)^2 d\rho \quad (18)$$

Eq. (18) is a normalized one as it is directly derived from Eq. (3). As for power calculation, a factor $cB_0^2/2\mu_0$ should be multiplied, where $B_0$ is the normalizing magnetic field for Eq. (3), $\mu_0$ is the permeability of vacuum.

Take a $CO_2$ laser with wavelength 10.6μm for example, to produce a 47 periods with $\lambda_u = 0.53mm$, $K_x = 0.5$ undulator, it needs laser power 9.3TW. The laser beam parameters are listed in Table 1.

Table 1. BLU parameters

| Laser wavelength | 10.6μm |
|---|---|
| $k_\perp/k$ | 0.199 |
| $k_\parallel/k$ | 0.98 |
| M | 2 |
| $\lambda_u$ | 0.53mm |
| R | 5mm |
| L | 25mm |
| Periods | 47 |
| Laser Power | 9.3TW |
| $K_x(K)$ | 0.5 (0.707) |

## 5. Beam tracking

Here we take Shanghai soft X-ray linac beam [25] as an example for tracking, whose electron energy is 840MeV. With BLU parameters listed in Table 1, a 1.2Å hard X-ray radiation will be produced.

To get a maximum deflection force, it is nature to align the electron beam at the place where $J_{M+1}(k_\perp \rho)$ gets peak value. The amplitude of the deflecting force at where has a small range of flat top in radial direction which allows a relatively large part of radiations from an electron beam could coherently superimposed. In the tracking, both transverse and longitudinal parts of the EM fields are counted. The longitudinal electrical field causes an energy modulation of the electron beam, resulting a deflecting effect in radial direction. The defecting force is in axial symmetry, so the focus (for left handed helix) or defocus (for right handed helix) effects are in all transverse directions as shown in Fig.2.

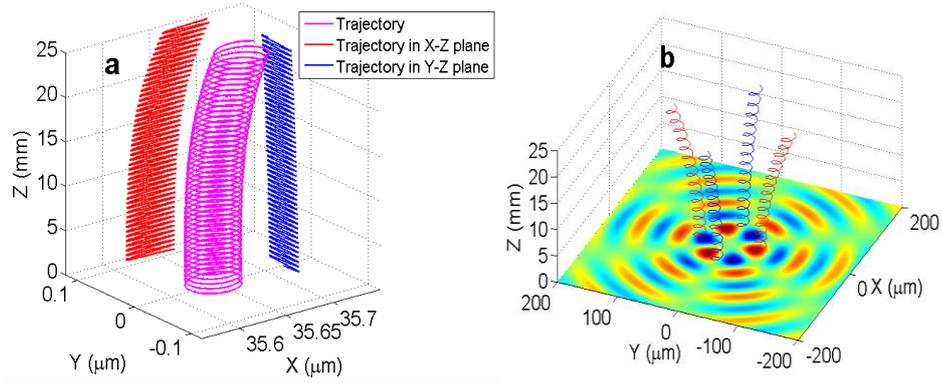

Figure 2. (a). Tracking result of electron trajectory. (b) Sketch of defocus effect, the beam trajectory is zoomed in to have a more clear view, at the bottom is the electrical field rotates anticlockwise.

To evaluate how many radiations from the electrons in transverse phase space could coherently superimposed, the radiation phase error is calculated as shown in Figure 3. The result shows BLU is sensitive to the transverse place where electron passes through however tolerant to slope angle of the electron.

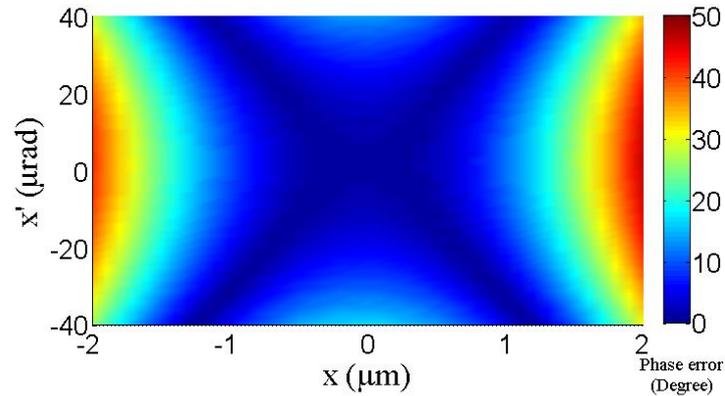

Figure 3. Radiation phase errors for electrons with different initial x-x' conditions.

## 6. Discussions

The BLU gets some unique properties that the other types of undulators do not hold. It only undulates a part of the electrons when the laser pulse is shorter than the electron beam. Because the group velocity of the Bessel light beam is at the speed of light, the interaction is limited at the place where electrons and the laser overlaps in longitudinal coordinate. This property can be used for short pulse (femtosecond) X-ray production when electron beam is long (picosecond).

The undulated electrons may shift backward in longitudinal coordinate, as the interaction is located in a part of longitudinal area in the electron beam which may cause an electron beam density modulation.

From the preliminary tracking results, it can be found out that BLU may radially focus (or defocus) electrons, which is different to quadrupoles and solenoids. The applications of those effects is under further investigations.

ACKNOWLEDGMENTS
The authors greatly thanks Prof. Alex Chao for useful discussions. The thanks also goes to Prof. Qiaogen Zhou for discussion on undulator performances and to Dr. Jianhui Chen for providing useful references.